\documentclass[aps,prl,showpacs,twocolumn,superscriptaddress,amsmath,amssymb,floatfix, longbibliography]{revtex4-1}
\usepackage{graphicx}
\usepackage{float}
\usepackage{latexsym,amsmath,amssymb,bm,euscript}
\usepackage{color}
\usepackage{wasysym}
\usepackage{epstopdf}
\usepackage[colorlinks=true,linkcolor=blue,citecolor=blue]{hyperref}
\usepackage{type1cm}
\usepackage{appendix}
\usepackage{mathtools}

\usepackage{amssymb}
\usepackage{stackengine}
\usepackage{scalerel}
\usepackage{xcolor}
\usepackage{graphicx}

\hypersetup{
           breaklinks=true,   
           colorlinks=true,   
           pdfusetitle=true,  
        }

\usepackage{tikz}

\DeclareMathOperator{\arccosh}{arccosh}

\newcommand\ssquare[1][.5]{\mathbin{\vcenter{\hbox{\scalebox{#1}{$\Square$}}}}}

\newcommand\sblacksquare[1][.5]{\mathbin{\vcenter{\hbox{\scalebox{#1}{$\blacksquare$}}}}}

\newcommand\scircle[1][.5]{\mathbin{\vcenter{\hbox{\scalebox{#1}{$\Circle$}}}}}

\newcommand\sbullet[1][.5]{\mathbin{\vcenter{\hbox{\scalebox{#1}{$\bullet$}}}}}

\newcommand\sdelta[1][.5]{\mathbin{\vcenter{\hbox{\scalebox{#1}{$\Delta$}}}}}

\newcommand\sblacktriangle[1][.5]{\mathbin{\vcenter{\hbox{\scalebox{#1}{$\blacktriangle$}}}}}

\newcommand\sdiamond[1][.5]{\mathbin{\vcenter{\hbox{\scalebox{#1}{$\lozenge$}}}}}

\newcommand\sblackdiamond[1][.5]{\mathbin{\vcenter{\hbox{\scalebox{#1}{$\blacklozenge$}}}}}

\newcommand\sDiamond[1][.5]{\mathbin{\vcenter{\hbox{\scalebox{#1}{$\Diamond$}}}}}

\usepackage{xcolor}
    \definecolor{myred}{HTML}{E07A60}
    \definecolor{myblue}{HTML}{687AB1}
    \definecolor{mygreen}{HTML}{66b063}

\begin{document}

\title{Negative superinflating bipartite fluctuations near exceptional points in $\mathcal{PT}$-symmetric models}

\author{Wei Pan}
\affiliation{Beijing Computational Science Research Center, Beijing 100084, China}

\author{Xiaoqun Wang}
\email[]{xiaoqunwang@zju.edu.cn}
\affiliation{School of Physics, Zhejiang University, Hangzhou, 310058, China}

\author{Haiqing Lin}
\email[]{haiqing0@csrc.ac.cn}
\affiliation{Beijing Computational Science Research Center, Beijing 100084, China}
\affiliation{School of Physics, Zhejiang University, Hangzhou, 310058, China}

\author{Shijie Hu}
\email[]{shijiehu@csrc.ac.cn}
\affiliation{Beijing Computational Science Research Center, Beijing 100084, China}
\affiliation{Department of Physics, Beijing Normal University, Beijing, 100875, China}

\begin{abstract}
We investigate bipartite particle number fluctuations near the rank-$2$ exceptional points (EPs) of $\mathcal{PT}$-symmetric Su-Schrieffer-Heeger models.
Beyond a conformal field theory of massless fermions, fluctuations or equivalently compressibility is negative definite and exhibits superinflation in leading order at EPs, due to the defectiveness in the biorthogonal Hilbert space.
Associated with the bipartite von Neumann entanglement entropy, a parameter in an anomalous correspondence referencing from a purely non-Hermitian limit helps characterize two inequivalent EP sets.
Our work paves the way for understanding the singularity of fluctuations relevant to EPs, more promisingly detectable in experiments.
\end{abstract}

\maketitle

{\it Introduction}.
Recently, emerged non-Hermitian quantum systems often destroy the reality of energy~\cite{El_Ganainy_2018} and give rise to distinct topological phases~\cite{Leykam_2017, Xu_2017, Gong_2018, Kawabata_2019}, an unconventional bulk-boundary correspondence~\cite{Lee_2016, Kunst_2018, Yao_2018, Lee_2019, Kunst_2019, Yokomizo_2019, Okuma_2019} defined in generalized Brillouin zones~\cite{Yao_2018}, a quasi-long-range order without breaking continuous symmetry~\cite{Lee_2014}, reactivated renormalization group flows~\cite{Ashida_2017, Nakagawa_2018}, a continuous phase transition without closure of the bulk gap~\cite{Matsumoto_2020}, and a non-Hermitian topological Anderson insulator~\cite{Lin_2022}.
Exceptional points (EPs)~\cite{Kato_1995}, caused by the (generalized) $\mathcal{PT}$ (parity $\cal P$ and time reversal $\cal T$) symmetry breaking, have been exploited for realizing unidirectional invisibility~\cite{Lin_2011, Regensburger_2012, Wang_2019}, single-mode lasers~\cite{Feng_2014, Hodaei_2014}, an EP-enhanced sensing~\cite{Chen_2017, Hodaei_2017}, the topological energy transfer~\cite{Xu_2016}, non-reciprocal wave propagation~\cite{R_ter_2010, Peng_2014} and so on~\cite{Ashida_2020}.

Unique to quantum systems~\cite{Furukawa_2009, Calabrese_2011}, nonlocal and nonlinear entanglement has evoked intense studies in recent decades, despite being difficult to measure experimentally~\cite{Cramer_2011, Levine_2011, Cardy_2011}.
To consider the necessary biorthogonal measurement due to the specification of probability rules in the {\it non-unitary} quantum mechanism~\cite{Brody_2013}, theorists have attempted to decorate the conformal field theory (CFT) for explaining unexpected scaling behaviors~\cite{Couvreur_2017, Bianchini_2014, Chang_2020, Narayan_2016, Bianchini_2014, Bianchini_2015, Dupic2018, Bianchini_2016, Jatkar_2017} as space is curved by non-Hermiticity~\cite{Lv_2022}.
At ordinary critical points (CPs), the effective central charge obtained by fitting the bipartite von Neumann entanglement entropy (BvNEE) is counted by the number of points on the complex energy topology cut by either the zero-real or imaginary axis~\cite{Guo_2021}, analogous to the {\it area law} in Hermitian counterparts~\cite{Eisert_2010}.
In contrast, EPs, representing the {\it defectiveness} of Hilbert space~\cite{Kato_1995}, lead to a continuously varying central charge~\cite{Lee_2022} or seemingly recover ordinary CPs~\cite{Chen_2022}, hardly corresponding to minimal models in CFT~\cite{Couvreur_2017}.

In the letter, we investigate the experimentally detectable bipartite particle number fluctuations (BPNFs)~\cite{Cramer_2011, Levine_2011, Cardy_2011} in $\cal PT$-symmetric Su-Schrieffer-Heeger (SSH) models, where the reality of both BvNEE and BPNFs is preserved at EPs, convenient for following discussions.
Confirmed by both analytical derivation and large-scale numerical calculations, our findings suggest several things:
1) BPNFs and the corresponding compressibility become negative definite in the presence of EPs in the thermodynamical limit (TDL), distinguished from ordinary CPs.
2) The leading-order BPNFs inflate as a square logarithmic function of the system size faster than the logarithmic correction in the area law of BvNEE.
3) Different from the Tomonago-Luttinger liquids (TLLs)~\cite{Song_2012}, a new correspondence between BvNEE and BPNFs gives a parameter for characterizing two sets of EPs.
Besides, the additional next-next-nearest-neighboring (NNNN) hoppings reveal more complicated inflating behaviors of BPNFs at EPs.

\begin{figure*}[t]
\begin{center}
\includegraphics[width=\linewidth]{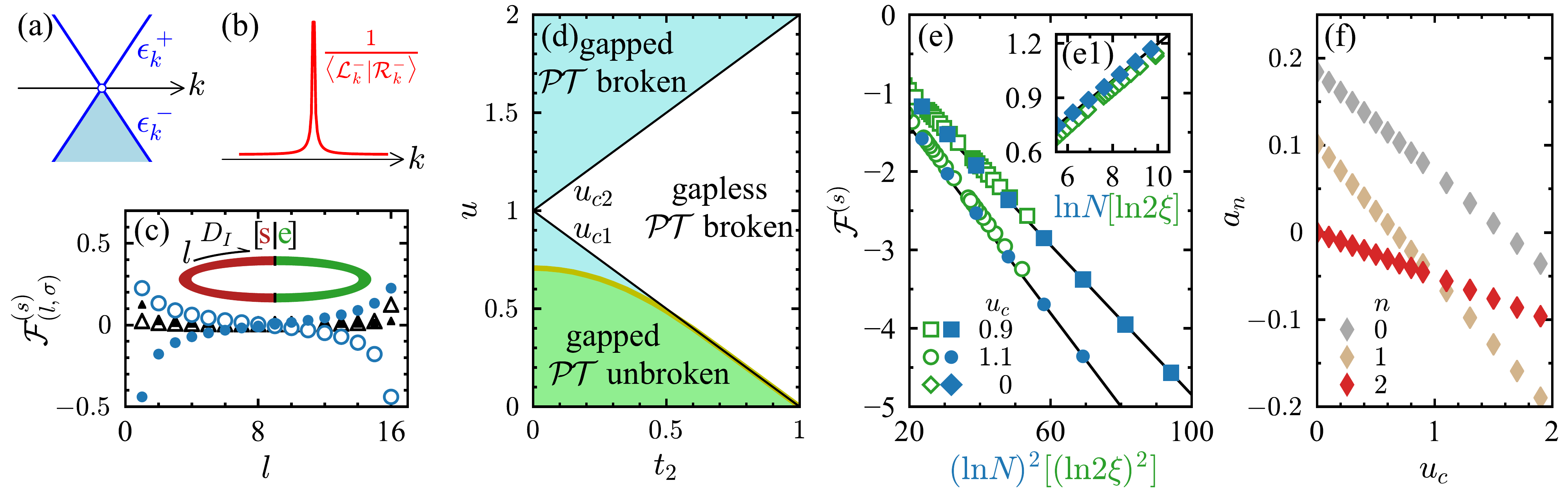}
\caption{
(a, b) The energy bands $\epsilon^{\pm}_{k}$ (blue lines) and the inverse of the norm $\langle {\cal L}^{-}_{k} \vert {\cal R}^{-}_{k} \rangle$ (red curves) in the vicinity of $k_c$ ({\color{blue} $\circ$}) at EP.
The low energy band is filled (blue shadow) at half-filling.
(c) The spatial distribution of partial fluctuations ${\cal F}^{(s)}_{(l,\sigma)}$ is shown for $u_c=0$ ({\color{black} $\sdelta[0.8]$} and {\color{black}$\sblacktriangle[0.8]$}) and $0.5$ ({\color{myblue} $\scircle[0.8]$} and {\color{myblue} $\sbullet[1.1]$}) when $N=64$.
Filled (open) symbols denote $\sigma=A$ ($B$).
Inset: a ring is partitioned into subsystem-$s$ (red) and $e$ (green) by two interfaces.
The site-$(l,\sigma)$ in the subsystem $s$ is $D_I$ unit cells away from the interface-$[s|e]$.
(d) The phase diagram of the model (\ref{modelham}) consists of gapped $\mathcal{PT}$-unbroken, gapless (blank) and gapped $\mathcal{PT}$-broken regions, separated by two EP sets $u_{c1}$ and $u_{c2}$ (black lines).
In gapped regions, real BPNFs ${\cal F}^{(s)}$ are either negative (cyan) or positive (green) in TDL.
(e) BPNFs ${\mathcal{F}}^{(s)}$ as a function of $(\ln N)^2$ (blue) or $(\ln 2\xi)^2$ (green) for $u_c = 0.9$ ({\color{mygreen} $\ssquare[1]$} and {\color{myblue} $\sblacksquare[0.8]$}) and $1.1$ ({\color{mygreen} $\scircle[0.8]$} and {\color{myblue} $\sbullet[1.1]$}).
Inset-(e1): it is plotted as a function of $\ln N$ (blue) and $\ln 2\xi$ (green) for $u_c = 0$ ({\color{mygreen} $\sdiamond[0.9]$} and {\color{myblue} $\sblackdiamond[0.9]$}).
Black lines indicate the best fittings. 
For convenience, we choose $0 \le u_{c} = u_{c1} \le 1$ and $1 \le u_{c} = u_{c2} \le 2$ for two EP sets, respectively.
(f) The coefficients $a_n$ ($n=0$, $1$ and $2$) in ${\cal F}^{(s)} = a_{2} (\ln N)^2 + a_{1} (\ln N) + a_{0}$ as a function of $u_{c}$.
}\label{fig1}
\end{center}
\end{figure*}

{\it Model}.
The non-Hermitian SSH model, in various forms, is a commonly selected platform for presenting rich physics~\cite{Lee_2016, Yao_2018, Yokomizo_2019}.
For a duplex lattice with sublattices $A$ and $B$, the $\mathcal{PT}$-symmetric one~\cite{Poli_2015, Zeuner_2015, Weimann_2016} reads
\begin{eqnarray}
\hat{H} &=& -\sum^{N}_{l=1} \left(t_1 \hat{c}_{l,A}^{\dagger} \hat{c}_{l,B} + t_2 \hat{c}_{l,B}^{\dagger} \hat{c}_{l+1,A} + {\rm h.c.}\right) \nonumber\\
&+& iu \sum^{N}_{l=1} \left(\hat{n}_{l,A} - \hat{n}_{l,B}\right),\label{modelham}
\end{eqnarray}
where ${\hat c}_{l,\sigma}$ (${\hat c}^{\dagger}_{l,\sigma}$) and ${\hat n}_{l,\sigma}={\hat c}^{\dagger}_{l,\sigma} {\hat c}_{l,\sigma}$ are the annihilation (creation) and particle number operators for fermions at site-$(l, \sigma)$ with $\sigma= A$ or $B$ (assigned values $y_{A}=-1$ and $y_{B}=0$),
$u \ge 0$ the equal-amplitude staggered potential originating from the standard gain-loss mechanism~\cite{Schomerus_2013},
$t_{1}$ and $t_{2}$ the hopping amplitudes with $1 = t_1 \ge t_2$ hereafter,
and $l$ runs over all $N$ unit cells.
The displacement for the site-$(l,\sigma)$ is $x_{(l,\sigma)}=2 l + y_{\sigma}$.

To study quantum criticality relevant to EPs at {\it half-filling}, we focus solely on the ground state with the eigenvalue having the smallest real and largest imaginary parts, considered the long-living steady state in open systems at zero temperature~\cite{Lee_2014, Matsumoto_2020, Liu_2021}.
In the phase diagram, the gapped $\mathcal{PT}$-unbroken, gapless and gapped $\mathcal{PT}$-broken regions are joined by two sets of EPs $u_{c1} = t_1 - t_2$ and $u_{c2} = t_1 + t_2$, which meet at $t_2=0$ in Fig.~\ref{fig1}(d).
In the main text, we mainly discuss $u_{c1}$ and summarize conclusions about $u_{c2}$ under the periodic boundary condition (PBC). Consistent results for general cases are discussed in the Supplemental Material~\cite{SM}.

{\it Correlation matrix}.
As $u_c=u_{c1}$, either a Hermitian CP $u_c = 0$ or a set of EPs $0 < u_c < 1$, the high ($+$) and low ($-$) energy bands $\epsilon^{\pm}_k = \pm \sqrt{|f_k|^2-u^2_c}$ with a structure factor $f_k = t_1 + t_2 e^{-ik}$, close the line-gap at the momentum $k_c=\pi$~\cite{Kawabata_2019}.
The correlation matrix ${\mathcal{C}}_{(l,\sigma),(l’,\sigma')} = \sum_k e^{ikD} {\mathcal{Q}}^{\sigma\sigma'}_k / N$ physically averages the distribution function ${\mathcal{Q}}^{\sigma\sigma'}_k = {\cal L}^{-*}_{k,\sigma} {\mathcal{R}}^{-}_{k,\sigma'} / \langle {\mathcal{L}}^{-}_{k} | {\mathcal{R}}^{-}_{k} \rangle$ over all momenta in the first Brillouin zone (FBZ).
$D=l'-l$ is the displacement in unit cells.
$|{\mathcal{L}}^{-}_{k} \rangle$ and $|{\mathcal{R}}^{-}_{k}\rangle$ are the left and right eigenvectors in the low energy band, respectively.
Under a particle-hole transformation for all sites and a rotation ${\mathbf{Z}}=\prod^{N}_{l=1} e^{i\pi {\hat n}_{l,B}}$ for all sublattices B, ${\hat H}$ turns into ${\hat H}^{\dagger}$ plus an irrelevant constant, which yields a symmetry {\rm sym}-${\mathcal{C}}$: ${\mathcal{C}} = {\mathbf{Z}}^{\dagger} (\mathbb{I} - {\mathcal{C}}^{\dagger}) {\mathbf{Z}}$.

For a momentum $k=k_c + \delta k$ very close to $k_c$, ${\cal Q}^{AA}_{k} = 1/2 + i {\cal Q}^{(nH)}_{k}$, ${\cal Q}^{BB}_{k} = 1/2 - i {\cal Q}^{(nH)}_{k}$, ${\cal Q}^{AB}_{k} = {\cal Q}^{(H)}_{k} - {\cal Q}^{(nH)}_{k}$, and ${\cal Q}^{BA}_{k} = {\cal Q}^{(H)\dagger}_{k} - {\cal Q}^{(nH)}_{k}$~\cite{SM}.
On the one hand, a linear energy dispersion $\epsilon^{-}_k \approx -v |\delta k|$ of fermions with a wave velocity $v = \sqrt{t_1 t_2}$ in Fig.~\ref{fig1}(a) leads to a step function ${\cal Q}^{(H)}_k = -i \pi \alpha e^{i\delta k / 2} \text{Sgn}(\delta k)/2$ depending of the sign of $\delta k$, which is the same as one for TLLs at the Hermitian CP $u_c=0$.
On the other hand, $1/\langle {\cal L}^{-}_{k} | {\cal R}^{-}_{k} \rangle \approx 1/|\delta k|$ in Fig.~\ref{fig1}(b) gives an essential singularity due to the defectiveness of the Hilbert space at EPs, where two eigenstates coalesce in the complex plane under a variation of a physical parameter~\cite{Kato_1995}.
Moreover, it deforms fermions and results in an extra correction ${\cal Q}^{(nH)}_k = \pi \beta / |\delta k|$.
The limit $t_2=0$ with two flat bands of EPs, where only ${\cal Q}^{(nH)}_k$ survives, is called the {\it purely non-Hermitian point}.
Coefficients $\alpha = t_2 / (\pi v)$ and $\beta=u_c / (2\pi v)$ are derived too.

For a proper distance $1 \ll |D| \ll N/2$, we get the correlation matrices ${\cal C}_{(l,A),(l',A)} \approx \delta_{l,l'}/2 -i \gamma {\cal Z}_{l,l'}$, ${\cal C}_{(l,B),(l',B)} \approx \delta_{l,l'}/2 + i \gamma {\cal Z}_{l,l'}$, ${\cal C}_{(l,A),(l',B)} \approx \gamma {\cal Z}_{l,l'} + \gamma {\cal W}_{(l,A),(l',B)}$, and ${\cal C}_{(l,B),(l',A)} \approx \gamma {\cal Z}_{l,l'} - \gamma {\cal W}_{(l,B),(l',A)}$ consisting of two characteristic functions
\begin{eqnarray}
{\cal W}_{(l,\sigma),(l',\sigma')} = \frac{\alpha}{x} = \alpha {\cal H}_{x},\ {\cal Z}_{l,l'} = \beta \ln \left|\frac{2D}{N}\right| = \beta {\cal L}_{D} \label{cmatrix}
\end{eqnarray}
with $x = x_{(l',\sigma')} - x_{(l,\sigma)}$ and a gauge phase $\gamma = (-1)^D$ uniformly-gradient in space.
For deformed fermions, the logarithmic function ${\cal Z}_{l,l'}$ rescaled by $N/2$, decays much slower relatively and governs the long-distance behavior of the correlation matrix, despite that the harmonic function ${\cal W}_{(l,\sigma),(l',\sigma')}$ is dominantly-large for a short distance $x \sim 1$.
In addition, $\cal C$ also inherits the $\cal PT$ symmetry from the ground state, so ${\cal C}_{(l,\sigma),(l',\sigma')}$ can be transformed into ${\cal C}^{*}_{(N-l+1,{\bar \sigma})(N-l'+1,{\bar\sigma'})}$ under the $\cal P$ transformation, where $\bar\sigma$ (${\bar\sigma}'$) is the reversion of $\sigma$ ($\sigma'$).

{\it BPNFs}.
A ring can be divided into two subsystems with equal sizes $N_s=N_e=N/2$, say, $s$ and $e$, by the interface-$[s|e]$ and $[e|s]$ as shown in Fig.~\ref{fig1}(c).
For the site-$(l,\sigma)$ in subsystem-$s$, the partial fluctuations ${\cal F}^{(s/e)}_{(l,\sigma)} = \sum_{(l',\sigma') \in s/e} {\cal G}_{(l,\sigma),(l',\sigma')}$ sum the second-order momentum ${\cal G}_{(l,\sigma),(l',\sigma')} = \langle {\hat n}_{l,\sigma} {\hat n}_{l',\sigma'} \rangle - \langle {\hat n}_{l,\sigma} \rangle \langle {\hat n}_{l',\sigma'} \rangle$ in subsystems.
According to Wick's theorem~\cite{Herviou_2019, Chang_2020}, ${\cal G}_{(l,\sigma),(l',\sigma')} = -{\cal C}_{(l,\sigma), (l',\sigma')} {\cal C}_{(l',\sigma'), (l,\sigma)}$ tells us that the fluctuations ${\cal F}^{(s)}_{(l,\sigma)} = - {\cal F}^{(e)}_{(l,\sigma)}$ originate from all possible pairs of hoppings out of and sequentially back to the subsystem-$s$.
In Hermitian systems, it is semi-positive definite due to ${\cal C}_{(l,\sigma), (l',\sigma')} = {\cal C}^{*}_{(l',\sigma'), (l,\sigma)}$.
However, because of {\rm sym}-$\cal C$ in the model, ${\cal G}_{(l,\sigma),(l',\sigma')} = (-1)^{y_\sigma + y_{\sigma'}} |{\cal C}_{(l,\sigma),(l',\sigma')}|^2$ remains real but is alternating in space generally.
So the negative contribution from the purely imaginary off-diagonal elements in the correlation matrix $\cal C$ is allowed.
For the unit-cell-$l$ in subsystem-$s$, which is $D_I = N / 2 - l \ll N/4$ unit cells away from the interface-$[s|e]$,
the partial fluctuations ${\cal F}^{(s)}_{l} \approx (\alpha^{2}/2) ({\cal H}_{2 D_I + 1} + {\cal H}_{2 D_I -1}) - 2\alpha \beta {\cal H}_{2 D_I} ({\cal L}_{D_I} + 1)$ although both ${\cal F}^{(s)}_{(l,A)}$ and ${\cal F}^{(s)}_{(l,B)}$ contain higher-order terms $\pm (\alpha\beta/2){\cal L}^{2}_{2 D_I}$~\cite{SM}.
At the Hermitian CP $u_c=0$, the positive partial fluctuations ${\cal F}^{(s)}_{(l,\sigma)} \approx 1/(2\pi^2 D_I)$ once $|D_I| \gg 1$ ({\color{black} $\sdelta[0.9]$} and {\color{black}$\sblacktriangle[0.8]$}) in Fig.~\ref{fig1}(c).
In contrast, ${\cal F}^{(s)}_{(l,\sigma)}$ at EPs has a strong oscillation in space ({\color{myblue} $\scircle[1]$} and {\color{myblue} $\sbullet[1.1]$}) and becomes negative when $\sigma=A$.

BPNFs ${\cal F}^{(s)}$ sum the partial fluctuations ${\cal F}^{(s)}_l$ over all unit cells in subsystem-$s$ and give ${\cal F}^{(s)} \approx a_2 (\ln N)^2 + a_1 (\ln N) + a_0$ with the analytical prediction $a_2=\alpha\beta$ and $a_1=(\alpha^2 + \alpha \beta)$~\cite{SM}.
At the Hermitian CP $u_c=0$, ${\cal F}^{(s)} \sim (\ln N) / (\pi^2 K)$ with a Luttinger parameter $K=1$~\cite{Song_2012}, indicating non-interacting fermions.
At EPs, the leading-order term $(t_2 - t_1)(\ln N)^2 / (2\pi^2 t_1)$ is negative and inflates faster, called the {\it negative superinflation} (Fig.~\ref{fig1}(e)).
And thus, the growing behavior of BPNFs concerning the system size qualitatively changes in the presence of EPs and nonzero $a_2$ can be recognized as a detectable signal.
Coefficients $a_1$ and $a_0$ in the subleading terms run as a function of $\alpha^2$ and $\alpha \beta$, both of which are proportional to $t_2 / t_1$ (Fig.~\ref{fig1}(f)).
So do the other EPs $u_{c2}$ without $\cal PT$ symmetry breaking.

In Hermitian systems, we can kick fermions out of subsystem-$s$ by simply tuning on a tiny chemical potential $\mu < 0$ in an extra Hamiltonian ${\hat H}_{\text{squeeze}} = - \mu \sum_{(l,\sigma) \in s} {\hat n}_{l, \sigma}$, because of the positive compressibility, or equivalently BPNFs~\cite{Landau}.
At EPs in the model (\ref{modelham}), negative BPNFs bring us the reverse diffusion: The number of fermions in subsystem-$s$ grows after the squeezing is applied.

In the gapped $\cal PT$-unbroken region~\cite{Chang_2020}, a Lorentz-invariant dispersion $\epsilon^{+}_k \approx \sqrt{\Delta^2 + v^2 \delta k^2}$ near $k_c$ suggests an exponentially-decaying tail of $\cal C$ once $|D| \gg \xi$ with a finite correlation length $1/\xi=\arccosh[(t^{2}_{1}+t^{2}_{2}-u^2)/(2 t_1 t_2)]$.
The tail only provides a constant ${a'}_0$ independent of $\xi$ if $(N/2) \gg \xi$, while the core within the region $|D| < \xi$ gives $a_2 [\ln (2\xi)]^{2} + a_1 [\ln (2\xi)] + a_0$ to ${\cal F}^{(s)}$ as shown in Fig.~{\ref{fig1}(e)} either.
Therefore, in Fig.~{\ref{fig1}(d)}, the blue region with negative BPNFs becomes wider as $t_2$ gets smaller.

In principle, EPs continuously link a gapped commensurate phase to a gapless incommensurate phase, where two critical exponents $\nu=\nu'=1/2$ on both sides imply the universality of the Pokrovsky-Talapov (PT) type~\cite{SM}, usually appearing in crystal surfaces~\cite{Huse_1982} and quantum Rydberg atomic systems~\cite{Chepiga_2021}.
Nevertheless, it does not have a well-defined Luttinger parameter $K=1$ because of the following anomalous correspondence.

\begin{figure}[t]
\begin{center}
\includegraphics[width=\linewidth]{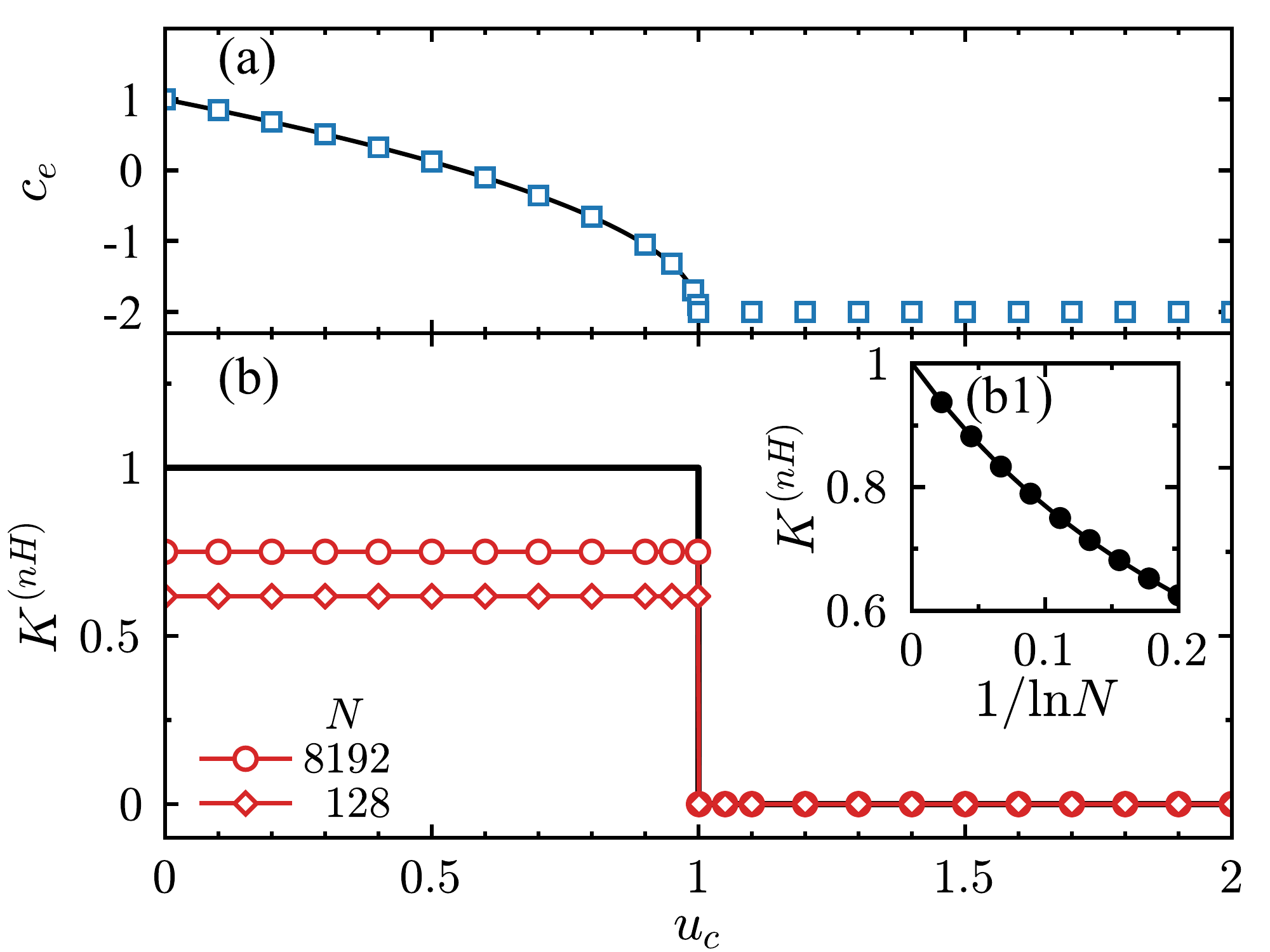}
\caption{
(a) The effective central charge $c_e$ ({\color{myblue} $\ssquare[1]$}) as a function of $u_c$, which matches the formula $c_e = 3\sqrt{1-u_c}-2$ (black curve) in the region $0 \le u_c \le 1$ very well.
(b) The parameter $K^{(nH)}$ as a function of $u_c$ for $N=128$ ({\color{myred} $\scircle[0.9]$}) and $8192$ ({\color{myred} $\sdiamond[0.7]$}).
The black line gives the analytical prediction in TDL.
Inset-(b1): The finite-size scaling of $K^{(nH)}$ ({\color{black} $\sbullet[1.2]$}) as a function of $1/ \ln N$ when $u_c = 0.5$.
}\label{fig2}
\end{center}
\end{figure}

{\it Anomalous entanglement-fluctuation correspondence}.
BvNEE ${\cal S} = -\text{Tr}(\rho_{s} \ln \rho_{s})$, where the reduced density matrix $\rho_{s} = \text{Tr}_{e} \rho$ for subsystem-$s$ is obtained by tracing out subsystem-$e$ in the non-Hermitian density matrix $\rho$.
Unlike Hermitian density matrices, non-Hermitian $\rho$ is useful for studying biorthogonal physics~\cite{Herviou_2019, Modak_2021}.
To suppose that $\cal C$ has a left eigenstate $|{\cal L}^{\cal C}\rangle$ and the right one $|{\cal R}^{\cal C}\rangle$ with an eigenvalue $w^{\cal C}$, it is easy to prove that $|{\cal L}^{\cal C}\rangle^{*}$ and $|{\cal R}^{\cal C}\rangle^{*}$ describe another eigenstate with eigenvalue of $1-w^{{\cal C}*}$ according to sym-$\cal C$.
Thus spectrum levels $\{w^{\cal C}_{q}\}$ are arranged in pairs, which also preserve the reality of the BvNEE in the $\cal PT$-unbroken region.
In addition, a combination of the translational symmetry under PBC and inherited ${\cal PT}$ one from the ground state reduces to a ${\cal P}_{s} {\cal T}_{s}$ symmetry for subsystem-$s$, which says ${\cal C}_{(l,\sigma),(l',\sigma')} = {\cal C}^{*}_{(N_{s} - l + 1,{\bar\sigma}),(N_{s} - l' + 1,{\bar\sigma'})}$.
For finite $N$, a transition caused by the ${\cal P}_{s}{\cal T}_{s}$ symmetry breaking near EPs brings us emerged finite-size effects, i.e., the concave form of the spatial profile of vNEE~\cite{Chang_2020, SM}.

For fixed $u$ close to EPs, as gradually growing $(N/2) \gg \xi$, $\cal S$ collapses to a logarithmic correction ${\cal S} \sim (c_{e}/3) \ln\xi$ in the area law~\cite{SM}.
The effective central charge $c_e = 1$ at $u_c=0$ depicts a CFT of massless fermions, while $c_e=-2$ at $u_c=t_1$ is relevant to a pure exceptional bound (EB) state~\cite{Lee_2022}.
In between two limits, the deformed fermions have a continuously varying $c_{e} = 3\sqrt{1 - u_c} - 2$ in Fig.~\ref{fig2}(a).

\begin{figure}[t]
\begin{center}
\includegraphics[width=\linewidth]{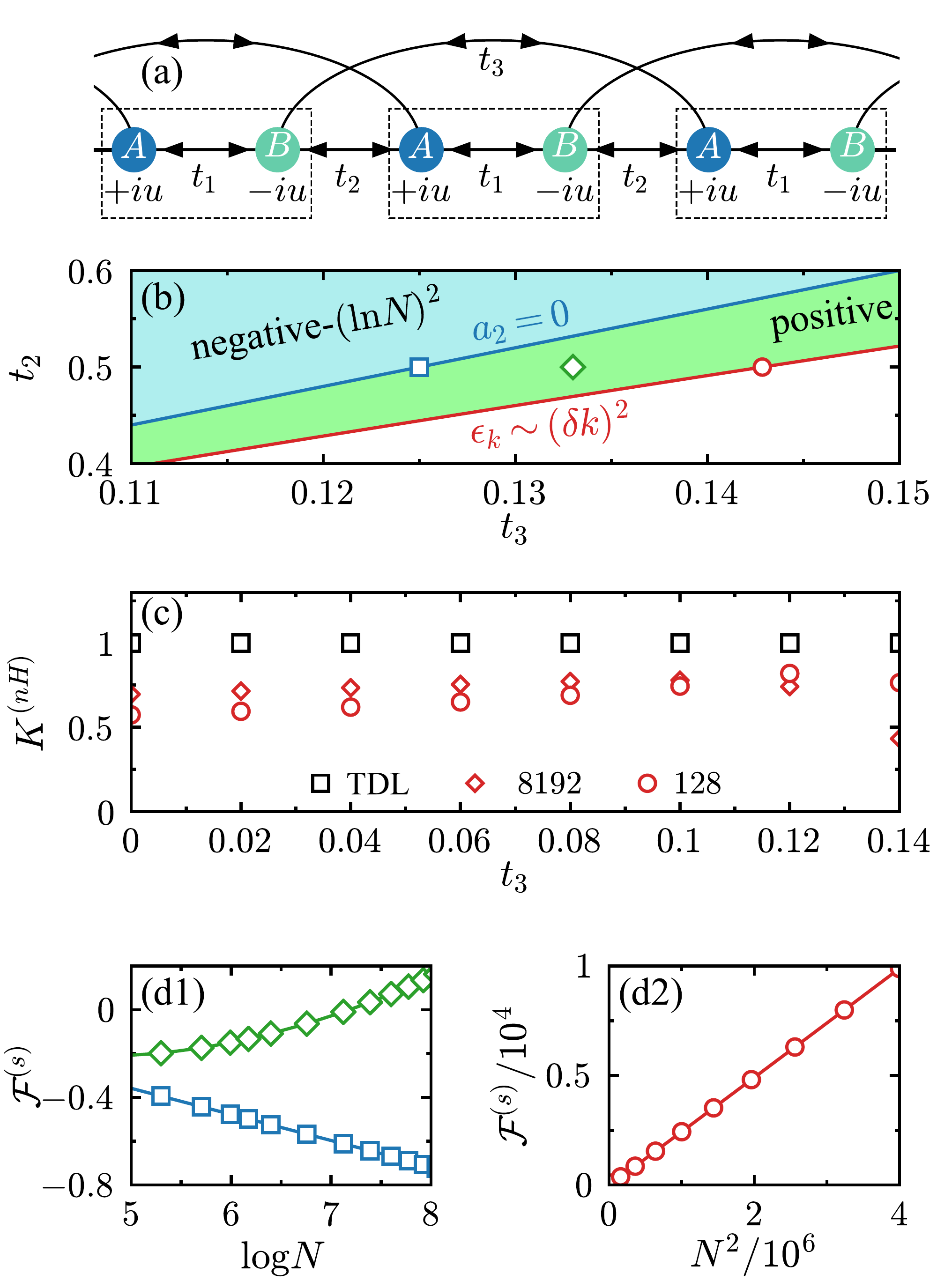}
\caption{
(a) The schematic picture of the model with NNNN hoppings $t_3$.
(b) The sign diagram of BPNFs at EPs $u_c = t_1 - t_2 + t_3$.
Adjacent to the negative-$(\ln N)^2$ (blue) region, the positive (green) one is sandwiched with two border lines $t^{L}_3=t_2/4$ (blue) and $t^{H}_3=t_1 t_2/(4t_1-t_2)$ (red).
In the blank region, none of the EPs $u_c$ can be found.
(c) The parameter $K^{(nH)}$ as a function of $t_3$ for $N=128$ ({\color{myred} $\scircle[0.9]$}), $8192$ ({\color{myred} $\sdiamond[0.7]$}), and TDL ({\color{black} $\ssquare[1]$}).
(d1, d2) Scaling behaviors of BPNFs at three typical parameters $t_3 = t^{L}_3$ ({\color{myblue} $\ssquare[1]$}), $0.133$ ({\color{mygreen} $\sDiamond[1]$}) and $t^{H}_3$ ({\color{myred} $\scircle[0.9]$}), also marked in (b).
In both (c), (d1), and (d2), we set $t_2=0.5$.
}\label{fig3}
\end{center}
\end{figure}

For TLLs, such as massless fermions at $u_c=0$ in the model (\ref{modelham}), BvNEE behaves simply as a linear function of BPNFs because they are both proportional to $\ln N$ (or $\ln \xi$).
Their corresponding relation yields a meaningful Luttinger parameter $K=3{\cal S}/(\pi^2 {\cal F}^{(s)})$ in TDL~\cite{Song_2012}, uniquely determining the conformal tower structure in CFT and giving the low energy description of quantum field theory together with the wave velocity~\cite{Giamarchi_2003}.

However, due to the attached singularity at EPs, the superinflation of BPNFs compared to the regular inflation of BvNEE implies that most fluctuations do not create entanglement between two subsystems~\cite{Modi_2012}.
It also results in $K=0$ in TDL and the invalidness of the simple correspondence, which is not perturbable from the Hermitian CP $u_c=0$.
By contrast, referencing from the purely non-Hermitian point $t_2=0$, we define a new parameter $K^{(nH)} = (\delta {\cal S})^2 / (2\pi^2 \delta {\cal F}^{(s)})$, where $\delta {\cal S}$ and $\delta {\cal F}^{(s)}$ denote the corresponding discrepancy of the BvNEE and BPNFs, respectively.
Because of the non-analytical form appearing in the zeroth-order BvNEE as $t_2$ goes to zero, only the first and second-order terms are involved in $\delta {\cal S}$ and $\delta {\cal F}^{(s)}$.
In Fig.~\ref{fig2}(b), We find that $K^{(nH)}=1$ for EPs $u_{c1}$ but $0$ for $u_{c2}$ in TDL.
For finite $N$, we still observe evident jumps at the transition point $u_c=1$.

{\it Sign of superinflating BPNFs}.
In the model (\ref{modelham}), the leading-order BPNFs at EPs remain negative, regardless of the signs of $t_1$, $t_2$, and $u$.
For instance, to set EPs $u_c = -u_{c1} < 0$ or equivalently apply a time-reversal transformation to the model (\ref{modelham}), BPNFs keep invariant.

In a revised model with NNNN hoppings $t_3$~\cite{Yao_2018} shown in Fig.~\ref{fig3}(a), the sign of BPNFs can be reversed, where a structure factor $f_k \approx u_c + i (t_2 - 2 t_3)\delta k$ and a low energy band $\epsilon^{-}_{k} \approx -v |\delta k|$ with a wave velocity $v = \sqrt{t_1t_2+t_2t_3 - 4t_1 t_3}$ at EP $u_c = t_1 - t_2 + t_3$.
Accordingly, ${\cal Q}^{(nH)}_k = \pi \beta / |\delta k|$ with $\beta = u_c/(2\pi v)$ keeps consistent in form with that of the model (\ref{modelham}).
Differently, ${\cal Q}^{(H)}_k = -i\pi (\alpha_1 e^{i\delta k/2} - \alpha_2 e^{i\delta k}) \text{Sgn}(\delta k) / 2$ with $\alpha_1 = t_2 / (\pi v)$ and $\alpha_2 = 2t_3 / (\pi v)$, which introduces an additional term ${\cal H}_{2 D + 2}$.
The coefficient in the leading-order BPNFs $a_2 = \beta (\alpha_1 - 2 \alpha_2)$ becomes positive once $t_3 > t_2 / 4 = t^{L}_3$ (Fig.~\ref{fig3}(b)).
Besides, two things are concerned.
Firstly, due to the competition between $\alpha_1$ and $\alpha_2$ terms, the superinflation of BPNFs reduces to a regular scaling of ${\cal F}^{(s)} \propto \ln N$ at $t^{L}_3$ in despite of a finite non-Hermitian $u_c = t_1 - 3 t_2 /4 > 0$ (Fig.~\ref{fig3}(d1)).
Secondly, in the zero-wave-velocity limit of $t^{H}_3 = t_1 t_2 / (4t_1 - t_2)$, the dispersion $\epsilon^{-}_k\approx v^{(2)} (\delta k)^2$ with $v^{(2)} = t_1 \sqrt{t_2 / (4t_1-t_2)}$ becomes parabolic in leading order, and BPNFs inflate faster, i.e., ${\cal F}^{(s)} \propto N^2$ in Fig.~\ref{fig3}(d2).

Upon examining the entanglement-fluctuation correspondence, it becomes evident that all EPs related to the $\cal PT$-symmetry breaking exhibit a consistent character in non-Hermitian SSH models.
For fixed $t_2=0.5$ as shown in Fig.~\ref{fig3}(c), referencing from the pure non-Hermitian point $t_2=t_3=0$, $K^{(nH)} = 1$ always in TDL, although NNNN hoppings $t_3 > 0$ affect both the leading-order BPNFs and BvNEE.
The theory for relatively complex finite-size effects has yet to be determined.

{\it Summary and discussion}.
We unveiled the superinflation of bipartite particle number fluctuations at exceptional points in $\cal PT$-symmetric SSH models.
The non-Hermitian correlations yield negative compressibility and reverse diffusion in broad parameter regimes in the thermodynamical limit.
The ratio coefficient in an anomalous entanglement-fluctuation correspondence helps characterize two EP sets.
Unlike entanglement entropy, behaviors of particle number fluctuations are more promisingly detectable in recent~\cite{Liang_2022} or future experiments.

Thanks to recent advances in non-Bloch band theory, it is now feasible to characterize a range of fascinating non-Hermitian phenomena, e.g., non-Hermitian skin effects~\cite{Yao_2018, Kunst_2018}, by employing the (auxiliary) generalized Brillouin zone (GBZ)~\cite{Yao_2018, Yang_2020}.
However, on condition that the collective behavior of fermions becomes dominant, the validity of the single-particle scenario needs to be further investigated, hinted by adverse examples mentioned in some recent works~\cite{Lee_2020, Mu_2020}.
We present a new perspective on examining the criticality of exceptional points in non-Hermitian quantum systems through the entanglement-fluctuation correspondence.

\begin{acknowledgments}
We thank Po-Yao Chang and Bo Yan for the grateful discussion.
This work is supported by the National Program on Key Research Project (Grant No.~MOST2022YFA1402700).
W.~P. acknowledges funding from Beijing Computational Science Research Center.
S.~H. acknowledges funding from the National Science Foundation of China under Grants No.~12174020.
H.~L. acknowledges funding from the National Science Foundation of China under Grants No.~12088101.
S.~H. and H.~L. further acknowledge support from Grant NSAF U1930402.
We acknowledge computational resources from Tianhe-2JK at the Beijing Computational Science Research Center (CSRC) and Quantum Many-body {\rm I} cluster at the School of Physics and Astronomy, Shanghai Jiaotong University.
\end{acknowledgments}

\bibliography{ref}
\end{document}


\title{\textbf{\large Supplemental Material for:\\ Negative superinflating bipartite fluctuations near exceptional points in $\cal PT$-symmetric models}}

\author{Wei Pan}
\affiliation{Beijing Computational Science Research Center, Beijing 100084, China}

\author{Xiaoqun Wang}
\email[]{xiaoqunwang@zju.edu.cn}
\affiliation{School of Physics, Zhejiang University, Hangzhou, 310058, China}

\author{Haiqing Lin}
\email[]{haiqing0@csrc.ac.cn}
\affiliation{Beijing Computational Science Research Center, Beijing 100084, China}
\affiliation{School of Physics, Zhejiang University, Hangzhou, 310058, China}

\author{Shijie Hu}
\email[]{shijiehu@csrc.ac.cn}
\affiliation{Beijing Computational Science Research Center, Beijing 100084, China}
\affiliation{Department of Physics, Beijing Normal University, Beijing, 100875, China}

\begin{abstract}
The Supplemental Material provides comprehensive information on $\cal PT$-symmetric Su-Schrieffer-Heeger models.
Firstly, we briefly derive the correlation matrix, partial particle number fluctuations, leading-order bipartite particle number fluctuations, and correlation length.
Secondly, we discuss the intriguing commensurate-to-incommensurate transition at exceptional points, which may belong to a universality class of non-Hermitian Pokrovsky-Talapov.
Thirdly, we demonstrate a novel spatial profile of the von Neumann entanglement entropy resulting from the ${\cal P}_{s} {\cal T}_{s}$ symmetry breaking.
Finally, we present additional data for the open boundary condition.
\end{abstract}

\maketitle

\renewcommand\thefigure{\thesection S\arabic{figure}}

\setcounter{figure}{0}

\section{I.~Correlation matrix}
This section explains how the correlation matrix $\cal C$ is derived.
The quadratic Hamiltonian of the Su-Schrieffer-Heeger (SSH) model presented in the main text sums the partial Hamiltonian ${\hat H}_k$ over all momenta $k$ in the first Brillouin zone (FBZ).
Unless otherwise stated, we typically use the periodic boundary condition (PBC), so $k=2m\pi/N$ and the integer number $m$ ranges from $0$ to $N-1$, where the chain is composed of $N$ unit cells.
For any $k$,
\begin{eqnarray}
{\hat H}_{k}=
\left(\begin{array}{cc}
  {\hat c}^{\dagger}_{k,A} & {\hat c}^{\dagger}_{k,B}
\end{array}\right)
\left(\begin{array}{cc}
       iu & -f_{k} \\
-f^{*}_{k} & -iu
\end{array}\right)
\left(\begin{array}{c}
     {\hat c}_{k,A}  \\
     {\hat c}_{k,B} 
\end{array}\right)
\end{eqnarray}
with the creation and annihilation operators
\begin{eqnarray}
{\hat c}_{k,\sigma} = \frac{1}{\sqrt{N}} \sum^{N}_{l=1} {\hat c}_{l,\sigma} e^{-i k l},\quad{\hat c}^{\dagger}_{k,\sigma} = \frac{1}{\sqrt{N}} \sum^{N}_{l=1} {\hat c}^{\dagger}_{l,\sigma} e^{i k l},
\end{eqnarray}
and the structure factor $f_k = t_1 + t_2 e^{-i k}$ defined as usual.

The model has high ($+$) and low ($-$) energy bands $\epsilon^{\pm}_k = \pm\sqrt{2t_1 t_2 \cos k + t^2_1 + t_2^2 - u^2}$, which open a gap at the center momentum $k_c = \pi$ in the gapped $\cal PT$-unbroken region $0 \le u < u_{c1} = t_1 - t_2$.
The gap closes at $u_c=u_{c1}$, corresponding to an ordinary critical point (CP) $u_c=0$ or a set of exceptional points (EPs) $0 < u_c < 1$.
Two-dimensional vectors $|{\cal L}^{-}_{k} \rangle = ({\cal L}^{-}_{k,A}, {\cal L}^{-}_{k,B})^{T}$ and $|{\cal R}^{-}_{k}\rangle = ({\cal R}^{-}_{k,A}, {\cal R}^{-}_{k,B})^{T}$ represent the corresponding left and right eigenvectors in the low energy band $\epsilon^{-}_{k}$, respectively.
Here, ${\cal L}^{-}_{k,A} = {\cal R}^{-*}_{k,B} = f_{k}$ and ${\cal L}^{-}_{k,B} = {\cal R}^{-}_{k,A} = p_k = - iu - \epsilon^{-}_{k}$.

In the correlation matrix
\begin{eqnarray}
{\cal C}_{(l,\sigma),(l',\sigma')} = \frac{1}{N} \sum_k e^{i k D} {\cal Q}^{\sigma\sigma'}_k,\label{CMAT}
\end{eqnarray}
with a displacement $D=l'-l$ in unit cells,
the momentum distribution function is
\begin{eqnarray}
{\cal Q}^{\sigma\sigma'}_k = {\cal L}^{-*}_{k,\sigma} {\cal R}^{-}_{k,\sigma'} / ({\cal L}^{-*}_{k,A} {\cal R}^{-}_{k,A} + {\cal L}^{-*}_{k,B} {\cal R}^{-}_{k,B}).
\end{eqnarray}
Particularly,
\begin{align}
\begin{split}
{\cal Q}^{AA}_{k} &= \frac{1}{2} + i\frac{u}{2\epsilon^{-}_{k}},\quad {\cal Q}^{AB}_{k} = -\frac{f^{*}_{k}}{2\epsilon^{-}_{k}},\\
{\cal Q}^{BB}_{k} &= \frac{1}{2} - i\frac{u}{2\epsilon^{-}_{k}},\quad {\cal Q}^{BA}_{k} = -\frac{f_{k}}{2\epsilon^{-}_{k}}.
\end{split}
\end{align}

For $k = k_c + \delta k$ very close to $k_c$ at $u_c$, we get $f_k = t_1 - t_2 e^{-i\delta k} \approx u_c + i t_2 \delta k$, $p_k \approx -i u_c + v |\delta k|$ and a wave velocity $v=\sqrt{t_1 t_2}$, up to the first order of $\delta k$.

Firstly, the momentum distribution functions for $AA$ and $BB$ become
\begin{align}
{\cal Q}^{AA}_{k} = \frac{1}{2} +i{\cal Q}^{(nH)}_{k},\quad {\cal Q}^{BB}_{k} = \frac{1}{2} -i{\cal Q}^{(nH)}_{k},
\end{align}
with a function ${\cal Q}^{(nH)}_k = \pi\beta/|\delta k|$ and $\beta=u_c / (2\pi v) = (\sqrt{t_2 / t_1} - \sqrt{t_1 / t_2})/2\pi$.
${\cal Q}^{(nH)}_k$ becomes singular at $\delta k = 0$.
The function of $1/|\delta k|$ originates from the loss of the biorthogonalization relation between the left and right eigenvectors at EPs $u_c > 0$ and $\beta \ne 0$, also called the {\it defectiveness} of the Hilbert space in the main text.
Meanwhile, we employ a trick of removing the ill-defined contributions at or near $k_c$ in the physical sum in Eq.~(\ref{CMAT}).
In Sec.~V, we also examine the negligible effects of the trick in the gapless $\cal PT$-broken region.

Secondly, the term for $AB$ is
\begin{eqnarray}
\begin{split}
{\cal Q}^{AB}_{k} &= \frac{1}{2v|\delta k|} \left[u_c + t_2e^{i\delta k/2}(e^{-i\delta k/2}-e^{i\delta k/2})\right]\\
&\approx \frac{1}{2v|\delta k|} \left[u_c -i t_2 \delta k e^{i\delta k/2}\right] = -{\cal Q}^{(nH)}_k + {\cal Q}^{(H)}_k,\quad
\end{split}
\end{eqnarray}
and similarly ${\cal Q}^{BA}_{k} \approx -{\cal Q}^{(nH)}_k + {\cal Q}^{(H)\dagger}_k$,
where we define a step function ${\cal Q}^{(H)}_k = -i \pi \alpha e^{i\delta k / 2} \text{Sgn}(\delta k)/2$, $\alpha = t_2 / (\pi v) = \sqrt{t_2 / t_1} / \pi$ and a conventional sign function ``$\text{Sgn}$".
It is noticed that the non-singular term $\text{Sgn}(\delta k)$ also exists for massless fermions at the Hermitian CP $u_c=0$.
Besides, ${\cal Q}^{AA}_{k}=({\cal Q}^{BB}_{k})^{*}$ and ${\cal Q}^{BA}_k = ({\cal Q}^{AB}_k)^{*}$ obviously due to the $\cal PT$-symmetric eigenvectors $|{\cal L}^{-}_k\rangle$ and $|{\cal R}^{-}_k\rangle$ in the gapped $\cal PT$-unbroken region.

In the continuum limit $1 \ll D \ll (N / 2)$, we get
\begin{eqnarray}
\begin{split}
{\cal C}_{(l,A),(l',A)} &= \frac{1}{2\pi}\int_{-\pi}^{\pi} e^{i k D} {\cal Q}^{AA}_{k} d (\delta k)\\
&=\frac{1}{2} \delta_{l,l'} - i \beta \gamma \int_{\pi/N}^{\pi} \frac{e^{i \delta k D} d (\delta k)}{|\delta k|}\\
&\approx \frac{1}{2} \delta_{l,l'} - i \beta \gamma \int_{\pi/N}^{\pi} \frac{e^{i \delta k D} d (\delta k)} {\sqrt{2(1-\cos\delta k)}}\\
&= \frac{1}{2} \delta_{l,l'} - i \gamma {\cal Z}_{l,l'}\label{CAA}
\end{split}
\end{eqnarray}
with a uniform gauge gradient $\gamma = (-1)^D$, a new quasi-long-range correlation ${\cal Z}_{l,l'} = \beta {\cal L}_D$ and a logarithmic function ${\cal L}_{z} = \ln |2z/N|$ rescaled by $N/2$.
Here we get rid of the momentum $\delta k = 0$ and use a formula~\cite{Lee_2022}
\begin{eqnarray}
\int_{\pi/N}^{\pi} \frac{e^{i z D}}{\sqrt{2(1-\cos z)}}\text{d} z \approx {\cal L}_D.
\end{eqnarray}
Moreover, the term $\delta_{l,l'}/2$ in Eq.~(\ref{CAA}) gives the fermion density at a site, consistent with the prior {\it half-filling}.
Likewise, ${\cal C}_{(l,B),(l',B)} \approx \delta_{l,l'}/2 + i\gamma {\cal Z}_{l,l'}$.

The correlation matrix for $AB$ is obtained by integrating all momenta in FBZ and takes the form
\begin{eqnarray}
\begin{split}
&\quad\ {\cal C}_{(l,A),(l',B)} = \frac{1}{2\pi}\int_{-\pi}^{\pi} e^{i k D} {\cal Q}^{AB}_{k} d (\delta k)\\
&= \gamma {\cal Z}_{l,l'} + \frac{1}{2\pi} \int_{-\pi}^{\pi} e^{i k D} {\cal Q}^{(H)}_{k} d (\delta k)\\
&= \gamma \left( {\cal Z}_{l,l'} + \frac{\alpha}{2} \int_{0}^{\pi} \sin[\delta k (2 D + 1)/2] d(\delta k)\right)\\
&= \gamma \left({\cal Z}_{l,l'} + {\cal W}_{(l,A),(l',B)}\right).
\end{split}
\end{eqnarray}
We quickly discover that ${\cal W}_{(l,\sigma),(l',\sigma')} = \alpha {\cal H}_{x}$ is proportional to a harmonic function ${\cal H}_{z} = 1/z$, which is a well-known correlation for massless fermions.
And, the prefactor $\alpha$ is subject to renormalization by non-Hermitian effects.
The displacement (in sites) of the site-$(l',\sigma')$ with respect to the site-$(l,\sigma)$ is given by $x = 2 D + y(\sigma') - y(\sigma)$, depending of $\sigma$ and $\sigma'=A$ or $B$.
In particular, $x=2 D + 1$ for the $AB$-term, while $x=2 D - 1$ for the $BA$-term.
In a similar vein, ${\cal C}_{(l,B),(l',A)}$ can be expressed as $\gamma \left({\cal Z}_{l,l'} + {\cal W}_{(l,B),(l',A)}\right)$.
Only a purely logarithmic term survives as the wave velocity approaches zero at $t_2=0$.

\section{II.~Partial particle number fluctuations}
It is convenient to specify the range of summation prior to conducting analytical calculations.
In the gapped $\cal PT$-unbroken region, the correlation matrix preserves a symmetry where ${\cal C}_{(l,\sigma),(l',\sigma')}={\cal C}^{*}_{(N-l+1,{\bar\sigma}),(N-l'+1,{\bar\sigma'})}$ under PBC, as mentioned in the main text.
Therefore, for fixed $l$, the correlation matrix ${\cal C}_{(l,\sigma),(l',\sigma')}$ decays as a function of the distance $D$ and has a zero-value node at $l'=l+N/2$.
The unit cells in subsystem-$e$ are naturally regrouped into two sets.
The first set, set-$[s|e]$, includes unit cells from $l'=N/2+1$ to $N/2+l-1$, which are closer to interface-$[s|e]$.
The second set, set-$[e|s]$, comprises the left unit cells in subsystem-$e$ near interface-$[e|s]$.
Since both sets have equal contributions to partial particle number fluctuations, we only need to consider the contribution from interface-$[s|e]$.
By doubling this value, we can determine the total amount.

As previously discussed, the unit-cell-$l'$ of subsystem-$e$ contributes $({\cal Z}_{l,l'})^2 - ({\cal W}_{(l,A),(l',B)} + {\cal Z}_{l,l'})^2$ to ${\cal F}^{(e)}_{(l,A)}$ for site-$(l,A)$ in subsystem-$s$, which is decomposed into a purely power-law term $-({\cal W}_{(l,A),(l',B)})^2$ and a mixed term $-2 {\cal W}_{(l,A),(l',B)} {\cal Z}_{l,l'}$.
Thus, the sum of the purely power-law terms from set-$[s|e]$ is equal to
\begin{eqnarray}
\begin{split}
{\cal P}_{(l,A)}&=-\sum^{N/2 + l - 1}_{l'=N/2 + 1} ({\cal W}_{(l,A),(l',B)})^2\\
&= -\alpha^2 \sum^{N/2 - 1}_{D = D_{I} + 1} {\cal H}^2_{2 D+1}\quad
\end{split}
\end{eqnarray}
with a displacement $D_{I}=N/2-l$ in unit cells with respect to interface-$[s|e]$.

According to the properties of the Riemann Zeta function $\zeta(x)$, we get
\begin{eqnarray}
\sum^{p}_{n=1} {\cal H}^2_{2n+1} = \frac{3}{4} \zeta(2) - 1 - \int^{\infty}_{0} \frac{z e^{-(2p + 1)z}}{e^{2z}-1} dz.\label{sum}
\end{eqnarray}
Once $p \gg 1$, Eq.~(\ref{sum}) approximates to $(3/4) \zeta(2) - 1 - (1/2){\cal H}_{2p+1}$.
Therefore,
\begin{eqnarray}
\begin{split}
{\cal P}_{(l,A)} &= \alpha^2 \sum^{D_{I}}_{D=1} {\cal H}^2_{2 D + 1} - \alpha^2 \sum^{N/2 - 1}_{D=1} {\cal H}^2_{2 D + 1}\\
&\approx \frac{\alpha^2}{2} \left( {\cal H}_{N-1} - {\cal H}_{2 D_{I} + 1}\right).
\end{split}
\end{eqnarray}
Meanwhile, if $N \gg 1$, ${\cal P}_{(l,A)} \approx -(\alpha^2/2) {\cal H}_{2 D_{I} + 1}$ shown in the main text.
For site-$(l,B)$, similarly if $N \gg 1$,
\begin{eqnarray}
{\cal P}_{(l,B)} = -\alpha^2 \sum^{N/2 - 1}_{D=D_{I} + 1} {\cal H}^2_{2D-1} \approx -\frac{\alpha^2}{2} {\cal H}_{2D_{I}-1}.
\end{eqnarray}

Secondly, if $D_{I} \ll N/2$, the mixed term 
\begin{eqnarray}
\begin{split}
{\cal M}_{(l,A)} &= -2 \sum^{N/2 + l - 1}_{l'=N/2 + 1} {\cal W}_{(l,A),(l',B)} {\cal Z}_{l,l'}\\
&= -2\alpha \beta \sum^{N/2 - 1}_{D = D_{I} + 1} {\cal H}_{2D+1} {\cal L}_D\\
&\approx -\alpha \beta \int^{1 - 2/N}_{2D_{I}/N} \frac{\ln z}{z}dz= \frac{\alpha \beta}{2} {\cal L}^2_{D_{I}}.
\end{split}
\end{eqnarray}
Combined with ${\cal M}_{(l,B)} = 2 \sum^{N/2 + l - 1}_{l'=N/2 + 1} {\cal W}_{(l,B),(l',A)} {\cal Z}_{l,l'}$, the sum is given by
\begin{eqnarray}
\begin{split}
&\quad\ {\cal M}_{(l,A)} \!+\! {\cal M}_{(l,B)}\\
& = 2 \alpha \beta \sum^{N/2 - 1}_{D = D_{I} + 1} \left({\cal H}_{2D-1} \!-\! {\cal H}_{2D+1}\right) {\cal L}_{2D}\\
&\approx \frac{2\alpha\beta}{N} \int^{(N - 2)/N}_{2D_{I}/N} \frac{\ln z}{z^2} dz \approx 2\alpha\beta {\cal H}_{2D_{I}} ({\cal L}_{D_{I}} + 1).\quad
\end{split}
\end{eqnarray}

So for unit-cell-$l$,
\begin{eqnarray}
\begin{split}
&\quad {\cal F}^{(s)}_{l} = {\cal F}^{(s)}_{(l,A)} + {\cal F}^{(s)}_{(l,B)} = -{\cal F}^{(e)}_{(l,A)} - {\cal F}^{(e)}_{(l,B)}\\
&\approx (\alpha^{2}/2) ({\cal H}_{2D_{I} + 1} + {\cal H}_{2D_{I} - 1}) - 2\alpha \beta {\cal H}_{2D_{I}} ({\cal L}_{D_{I}} + 1).\quad\quad
\end{split}
\end{eqnarray}

\section{III.~Leading-order bipartite particle number fluctuations}
The bipartite particle number fluctuations (BPNFs) at the ordinary CP $u_c=0$ with $\alpha=1/\pi$ and $\beta=0$ are given by
\begin{eqnarray}
\begin{split}
{\cal F}^{(s)} &= \alpha^2 \sum^{N/4}_{D_{I}=1} ({\cal H}_{2D_{I} + 1} + {\cal H}_{2D_{I} - 1})\\
&\approx \alpha^2 \int^{N/4}_{1} \frac{dz}{z} \approx \frac{\ln N}{\pi^2}.
\end{split}
\end{eqnarray}

At EPs $u_c = u_{c1} >0$, the leading-order BPNFs arise from the mixed term, namely
\begin{eqnarray}
\begin{split}
{\cal F}^{(s)} &= -4\alpha \beta \sum^{N/4}_{D_{I}=1} {\cal H}_{2 D_{I}} {\cal L}_{D_{I}}\\
&\approx -2\alpha \beta \int^{1/2}_{2/N} \frac{\ln z}{z} dz \approx \alpha \beta (\ln N)^2.
\end{split}
\end{eqnarray}

\section{IV.~Correlation length}
In this section, we derive the analytical form of the correlation length in the gapped $\cal PT$-unbroken region.
For the momentum near $k_c$,  denoted as $k=k_c + \delta k$ , the high energy band $\epsilon^{+}_{k}$ can be approximated by a Lorentz form $\sqrt{\Delta^2 + v^2 (\delta k)^2}$ with a gap $\Delta = \sqrt{(t_1-t_2)^2 - u^2}$ and a wave velocity $v=\sqrt{t_1 t_2}$.
The correlation matrix has an alternative form as follows:
\begin{align}
{\cal C}_{(l,\sigma),(l',\sigma')} = \sum_m \frac{\langle {\cal L}^{-} | {\hat c}_{l,\sigma}^{\dagger} | {\cal R}^{m} \rangle \langle {\cal L}^{m} | {\hat c}_{l',\sigma'} | {\cal R}^{-} \rangle}{\langle {\cal L}^{-} | {\cal R}^{-} \rangle}.
\end{align}
Here, $|{\cal L}^{m}\rangle$ and $|{\cal R}^{m}\rangle$ refer to the left and right eigenvectors of an intermediate excited state with one less fermion regarding the ground state $|{\cal L}^{-}\rangle=\prod_{k \in \text{FBZ}} |{\cal L}^{-}_{k}\rangle$ and $|{\cal R}^{-}\rangle=\prod_{k \in \text{FBZ}} |{\cal R}^{-}_{k}\rangle$, respectively.

Due to a finite gap $\Delta$ or equivalently a finite correlation length $\xi$ at $k_c$, we suppose that ${\cal C}_{(l,\sigma),(l',\sigma')} \sim \exp(-D/\xi)$ for a long distance, which means that the probability of creating a fermion at unit-cell-$l$ after removing a fermion at unit-cell-$l'$ is an exponentially-vanishing function of the distance $D$.
The above statement yields a scattering equation in the continuum limit, that is,
\begin{eqnarray}
\sqrt{\Delta^2 + v^2 (\delta k)^2} \exp(-D/\xi) = 0
\end{eqnarray}
with $\delta k = -i \partial / \partial D$.

Solving the above equation gives the reciprocal relation $\xi = v / \Delta$ between the gap $\Delta$ and correlation length $\xi$.
To consider higher-order corrections, the scattering equation becomes
\begin{eqnarray}
\sqrt{t_1^2 + t_2^2 - u^2 + 2 t_1 t_2 \cos(\delta k)} \exp(-D/\xi) = 0.
\end{eqnarray}
And we obtain a solution, i.e.,
\begin{eqnarray}
\frac{1}{\xi} = \text{arccosh} \left(\frac{t^{2}_{1} + t^{2}_{2} - u^2}{2 t_1 t_2}\right).\label{CORRLEN}
\end{eqnarray}
Subsequently, the effective wave velocity is
\begin{eqnarray}
v_{\text{eff}} = \Delta \times \xi.\label{V}
\end{eqnarray}

\begin{figure}
\includegraphics[width=\linewidth]{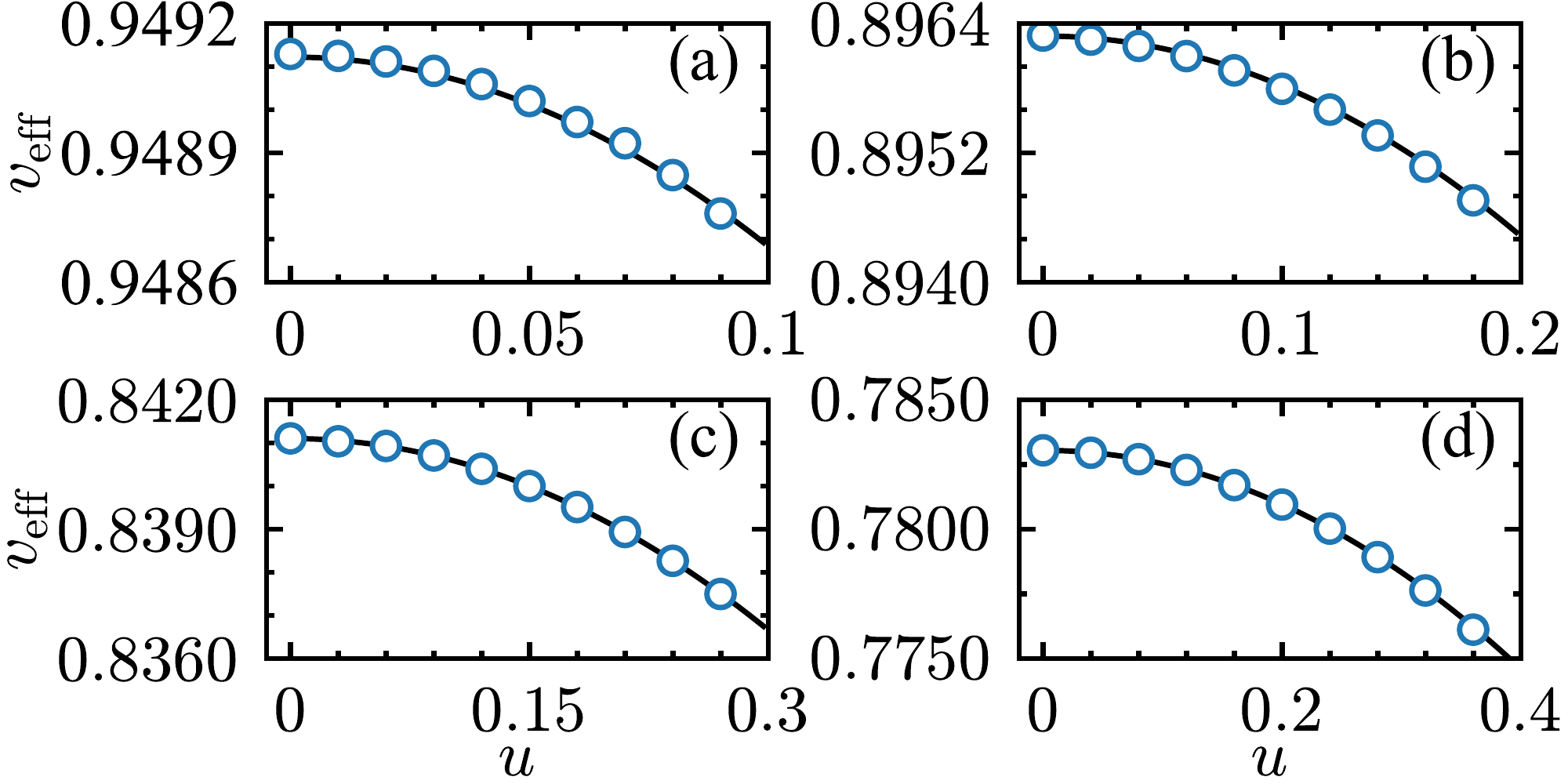}
\caption{The effective wave velocity $v_{\text{eff}}$ ({\color{myblue} $\scircle[0.8]$}) as a function of $u$ for different values of $t_2$ in the gapped $\cal PT$-unbroken region: $t_2=0.9$ (a), $0.8$ (b), $0.7$ (c), and $0.6$ (d)
The correlation length $\xi$ is obtained from the numerical fitting by Eq.~(\ref{XI_N}).
For instance, when $t_2=0.6$ and $u=0.2$, the fitting yields $\xi=2.25448$, $b_1=-0.50048$, $b_2=-1.00738$, $b_3=0.50088$, and $b_4=0.28158$, which is in close agreement with the analytical prediction $\xi=2.25444$ based on Eq.~(\ref{V}) (black lines).
We choose $t_1=1$ and $N=2048$ with PBC.
}\label{sfig1}
\end{figure}

Moreover, we numerically fit the correlation function $\ln|{\cal C}_{(l,A),(l',A)}|$ with higher-order corrections~\cite{Zauner_2015}
\begin{eqnarray}
\frac{-D}{\xi} + b_1\ln D + \ln\left(1 + \frac{b_2}{D} + \frac{b_3} {D^2} \right) + b_4.\quad\quad\label{XI_N}
\end{eqnarray}
In Fig.~\ref{sfig1}, we find that values of the effective wave velocity $v_{\text{eff}}$ keep consistent with the above-mentioned analytical prediction in Eq.~(\ref{V}).

\section{V.~Commensurate-to-incommensurate transition}
\begin{figure}
\includegraphics[width=\linewidth]{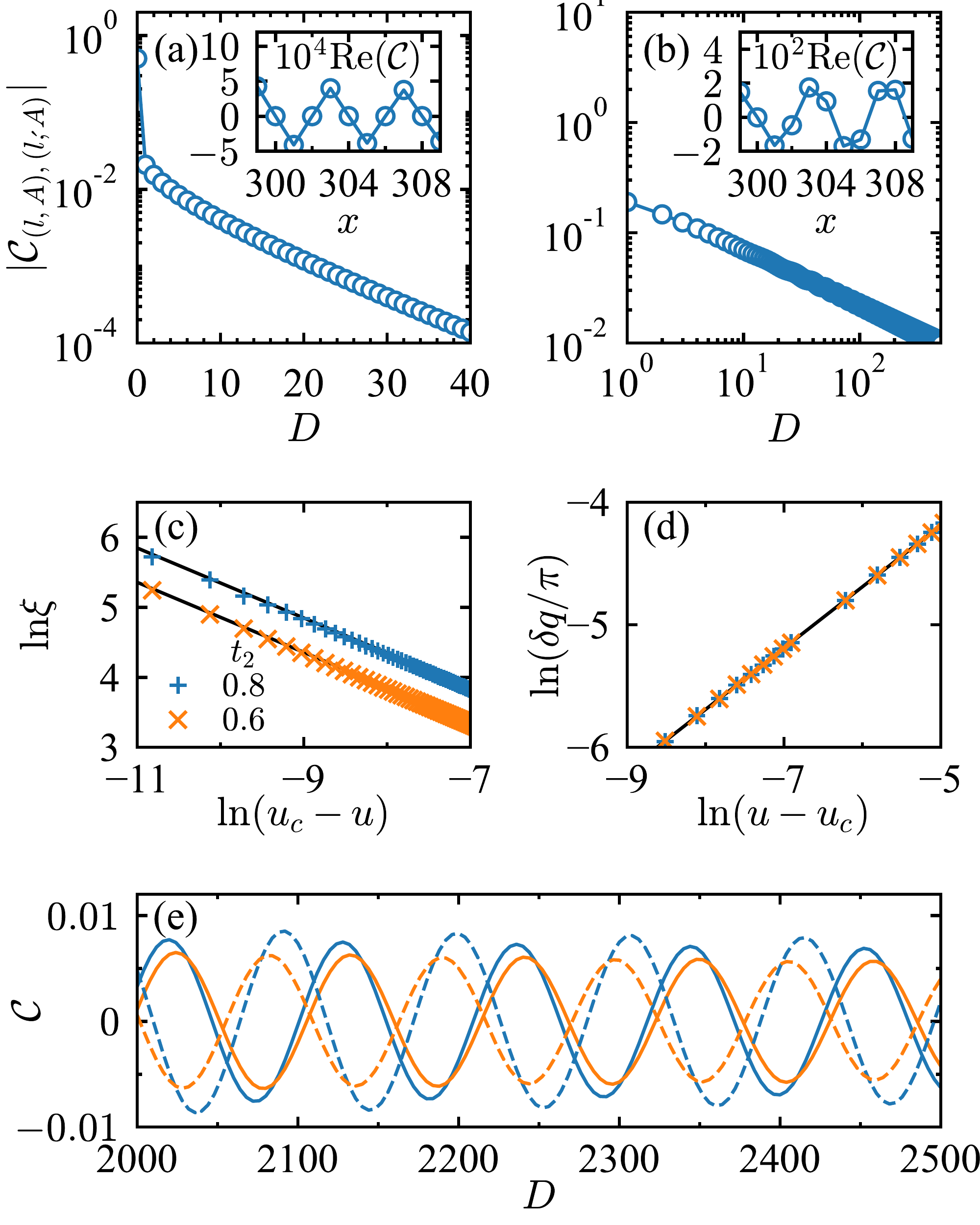}
\caption{
(a) and (b) Correlation matrix $|{\cal C}_{(l,A),(l',A)}|$ ({\color{myblue} $\scircle[0.8]$}) as a function of the distance $D$ for $t_2=0.6$.
(a) At $u=0.35$ in the gapped $\cal PT$-unbroken region, $|{\cal C}_{(l,A),(l',A)}|$ for $N=512$ decays exponentially for a long distance.
(b) At $u=0.45$ in the gapless $\cal PT$-broken region, $|{\cal C}_{(l,A),(l',A)}|$ behaves like a power-law function of $D$ for $N=65536$.
Insets show zoomed-in plots of the real part of $\cal C$ as a function of $x$, indicating a period of four sites in (a) and a non-integer period in (b), respectively.
Scaling behaviors of the correlation length $\xi$ (c) and the pitch angle discrepancy $\delta q$ (d) for $t_2=0.8$ ({\color{myblue} $\splus[0.8]$}) and $0.6$ ({\color{orange} $\scross[0.9]$}) with $N=16384$ match the analytical prediction in Eq.~(\ref{CORRLEN}) and (\ref{KCI}) (solid black lines) very well.
(e) The real (solid lines) and imaginary part (dashed lines) of the correlation matrix $\cal C$ at $u=0.41$ as a function of $D$ for $N=8192$ (blue), and $10808$ (orange).
We choose $t_1=1$ with PBC.
}\label{sfig2}
\end{figure}

In the case of the Hermitian model with $u=0$, two insulators can be distinguished by a berry phase $\pi$ detected in the filled low energy band, as well as a pair of zero-energy Majorana modes at edges~\cite{Su_1979}.
These insulators are connected at an ordinary CP $u_c=0$ with a CFT of massless fermions.
According to Eq.~(\ref{CORRLEN}), the correlation length $\xi$ approximates to $\sqrt{2} t_1 (t_1 - t_2)^{-\nu}$ with $\nu=1$ as $0 < 1 - t_2 / t_1 \ll 1$.

Starting from a Hermitian insulator as $t_1 > t_2$, a gap at $k_c$ remains open in the presence of a tiny imaginary staggered potential $u$.
The ground state retains the $\cal PT$ symmetry, and the correlation matrix $\cal C$ exhibits an oscillation with a commensurate period of two unit cells (or four lattice sites), as shown in the inset of Fig.~\ref{sfig2}(a).
As $u$ approaches the EP at $u_c > 0$, $1/\xi^2 \approx (t_1 - t_2) (u_c - u) / (2 t_1 t_2)$ or equivalently $\xi \approx \sqrt{2 t_1 t_2 / (t_1 - t_2)} (u_c - u)^{-\nu}$ with a critical exponent $\nu=1/2$.
This result is consistent with the closure of the gap $\Delta \approx (u_c - u)^{\nu}$ near the rank-$2$ EP.
The validity of the above conclusion is also supported by the large-scale numerical data in Fig.~\ref{sfig2}(c).

Once $u > u_c$, the presence of filled imaginary energy levels results in the breaking of $\cal PT$ symmetry.
In TDL, the closure of both real and imaginary gaps occurs at
\begin{eqnarray}
k^{I}_c = \arccos\left(\frac{u^2 - t_1^2 - t_2^2}{2 t_1 t_2}\right),\label{KCI}
\end{eqnarray}
which can be reached through subtly-tuned boundary conditions and the appropriate number of unit cells.
On the one hand, because of the gapless excitations at $k^{I}_c$, the up envelope of $\cal C$ gradually collapses to a power-law curve in double logarithmic scales.
On the other hand, a composite oscillation comprising a short period of four sites and a longer period structure suggests an incommensurate pitch angle $q = k^{I}_c/2 = k_c / 2 + \delta q$.
In TDL, the value of $\delta q$ is approximately given by $(u - u_c)^{\nu'}$, where another critical exponent $\nu'=1/2$, as depicted in Fig.~\ref{sfig2}(d).

In two-dimensional (2D) crystal surfaces, a transition from a gapped commensurate phase to a gapless incommensurate phase may exhibit the Pokrovsky-Talapov (PT) universality, characterized by critical exponents $\nu=\nu'=1/2$ on both sides of CP~\cite{Huse_1982}.
However, in Hermitian systems, robust evidence of the quantum counterpart of PT universality is limited due to the rare presence of quantum {\it floating phases}~\cite{Chepiga_2021}, which are incommensurate, gapless, and disordered.
Interestingly, in our model, EPs show a similar criticality of the PT universality.
The critical exponent $\nu=1/2$ governs the divergence of the correlation length near EPs in the gapped $\cal PT$-unbroken region.
While the other critical exponent $\nu'=1/2$ gives the scaling behavior of the pitch angle discrepancy $\delta q$ in the gapless $\cal PT$-broken region, which may be a non-Hermitian floating phase.

To avoid the numerical instability caused by the loss of the biorthogonalization relation between the left and right eigenvectors, we remove the contribution from the momentum $k^{I}_c$ and those nearby, which has a biorthogonalization error exceeding a specified criterion, such as $10^{-2}$ in practice.
Despite this adjustment, we found that the revised data still support consistent conclusions in TDL, similar to artificially adding a small momentum offset~\cite{Chang_2020}.
For instance, we take $t_2=0.6$ and $u=0.41$ with PBC as an example in Fig.~\ref{sfig2}(e) and compare the incommensurate pitch angle for different system sizes $N=8192$ and $10808$.
According to Eq.~(\ref{KCI}), we have $k^{I}_c=0.96300\pi$.
For $N=10808$, the mode with momentum $k=(10410/10808)\pi$ has a biorthogonalization error larger than the criterion, and we erase its contribution in the sum of Eq.~(\ref{CMAT}).
The best fitting of the correlation function suggests $q=0.48156\pi$.
In contrast, for $N=8192$, we consider all the low-energy levels, and the data fitting gives $q=0.48148\pi$.
Therefore, the relative error $0.0166\%$ caused by the trick is negligible.

\section{VI.~${\cal P}_s{\cal T}_s$ symmetry and the von Neumann entanglement entropy}
\begin{figure}
\includegraphics[width=\linewidth]{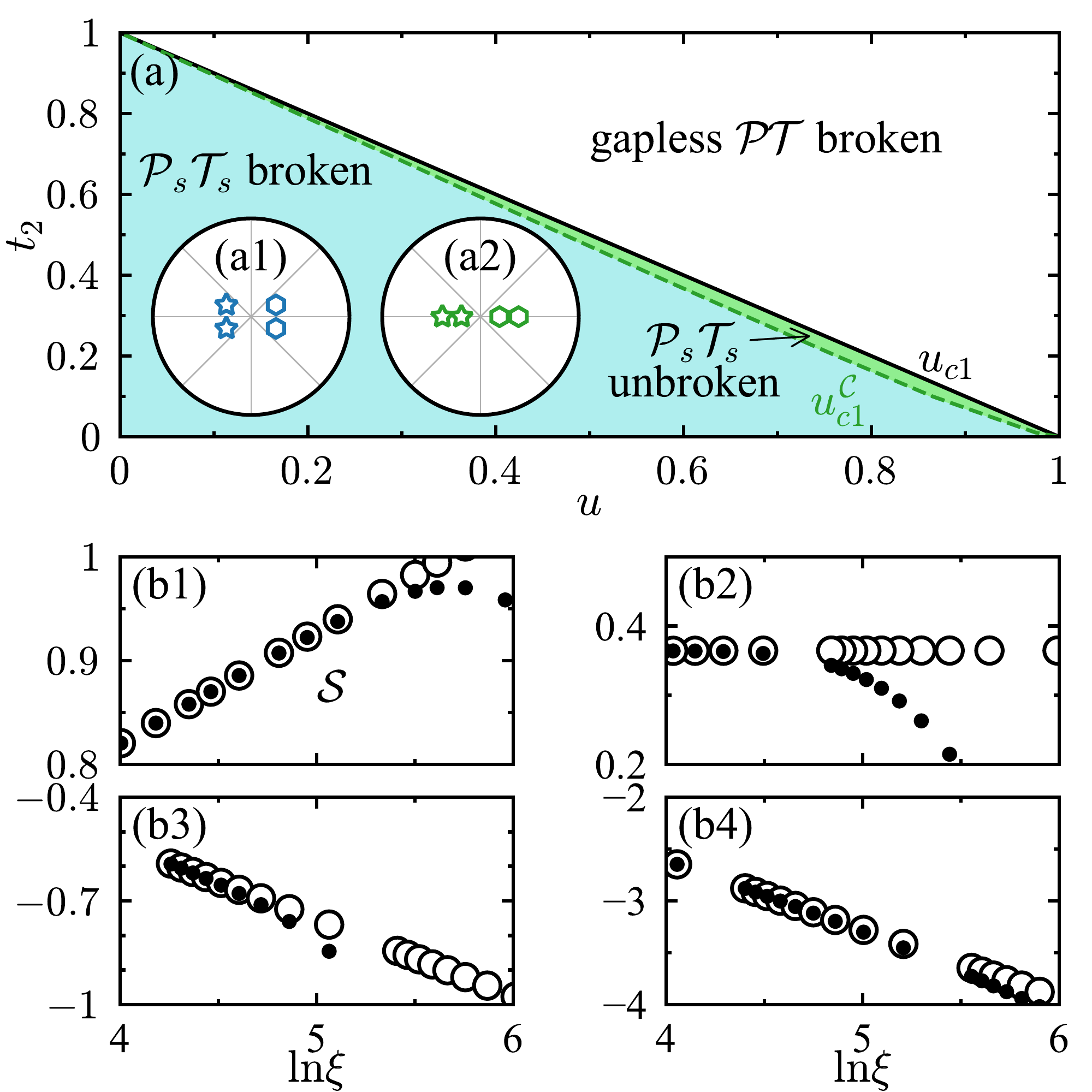}
\caption{
(a) The phase diagram of the model (1) in the main text.
The gapped $\cal PT$-broken (blank) and gapless $\cal PT$-unbroken regions are separated by a critical line $u_{c1}$ (black line).
For a finite $N=4$, the $\cal PT$-unbroken region is further divided into ${\cal P}_s {\cal T}_s$-broken (cyan) and ${\cal P}_s {\cal T}_s$-unbroken (green) regions by a quasi-critical line $u^{\cal C}_{c1}(N)$ (green dashed line).
(a1, a2) $2$D radar maps with a radius $r=|z|$ and an angle $\phi=\arg(z)$, for visualizing the spectrum $\{w^{\cal C}_{q}\}$ of the correlation matrix $\cal C$.
The set-$0$ $z=w^{\cal C}_{q}$ ({\color{myblue} $\openbigstar$} and {\color{mygreen} $\openbigstar$}) comprises all modes with $\text{Re}(w^{\cal C}_{q}) < 1/2$.
For the set-$1$ ({\color{myblue} $\varhexagon$} and {\color{mygreen} $\varhexagon$}) that includes the remaining modes, we assign $z=w^{\cal C}_{q} - 1$.
The display area is limited by a high bound $r=1$ (black outer circle border) and a low bound $r=10^{-2}$ (origin) in the log-$10$ scale.
(b1-b4) Von Neumann entanglement entropy $\cal S$ as a function of $\ln \xi$ for $t_2=0.6$ (b1), $t_2=4/9$ (b2), $t_2=0.2$ (b3) for EPs $u_{c1}$ and $t_2=0.5$ for $u_{c2}$ (b4).
$N=512$ ({\color{black} $\sbullet[1]$}) and $8192$ ({\color{black} $\scircle[1]$}) are compared.
Here, $t_1=1$ is chosen with PBC.
\label{sfig3}}
\end{figure}

Due to the ${\cal P}_s {\cal T}_s$ symmetry of the partial correlation matrix for subsystem-$s$, its spectrum levels are either real numbers or complex conjugated pairs~\cite{El_Ganainy_2018}.
Under PBC, the ${\cal P}_{s}{\cal T}_{s}$ symmetry breaks for an infinitesimal $u$, resulting in the emergency of complex conjugate pairs in the radar map Fig.~\ref{sfig3}(a1).
However, the ${\cal P}_{s} {\cal T}_{s}$ symmetry is restored once $u > u^{\cal C}_{c1}(N)$, characterized by a signal of two fresh real pairs in Fig.~\ref{sfig3}(a2), despite the stronger non-Hermitian $u$ near EPs.
In TDL, the ${\cal P}_{s}{\cal T}_{s}$ symmetry never survives in the whole $\cal PT$-unbroken region, i.e., $u^{\cal C}_{c1}(\infty)=u_{c1}$.
Consequently, for a fixed $u$ close to EPs, the bipartite von Neumann entanglement entropy (BvNEE) $\cal S$ follows an {\it area law} of ${\cal S} \sim (c_{e}/3) \ln\xi$ as $(N/2) \gg \xi$ in Fig.~\ref{sfig3}(b1-b4).

\section{VII.~Novel spatial profiles of the von Neumann entanglement entropy}
\begin{figure}
\includegraphics[width=\linewidth]{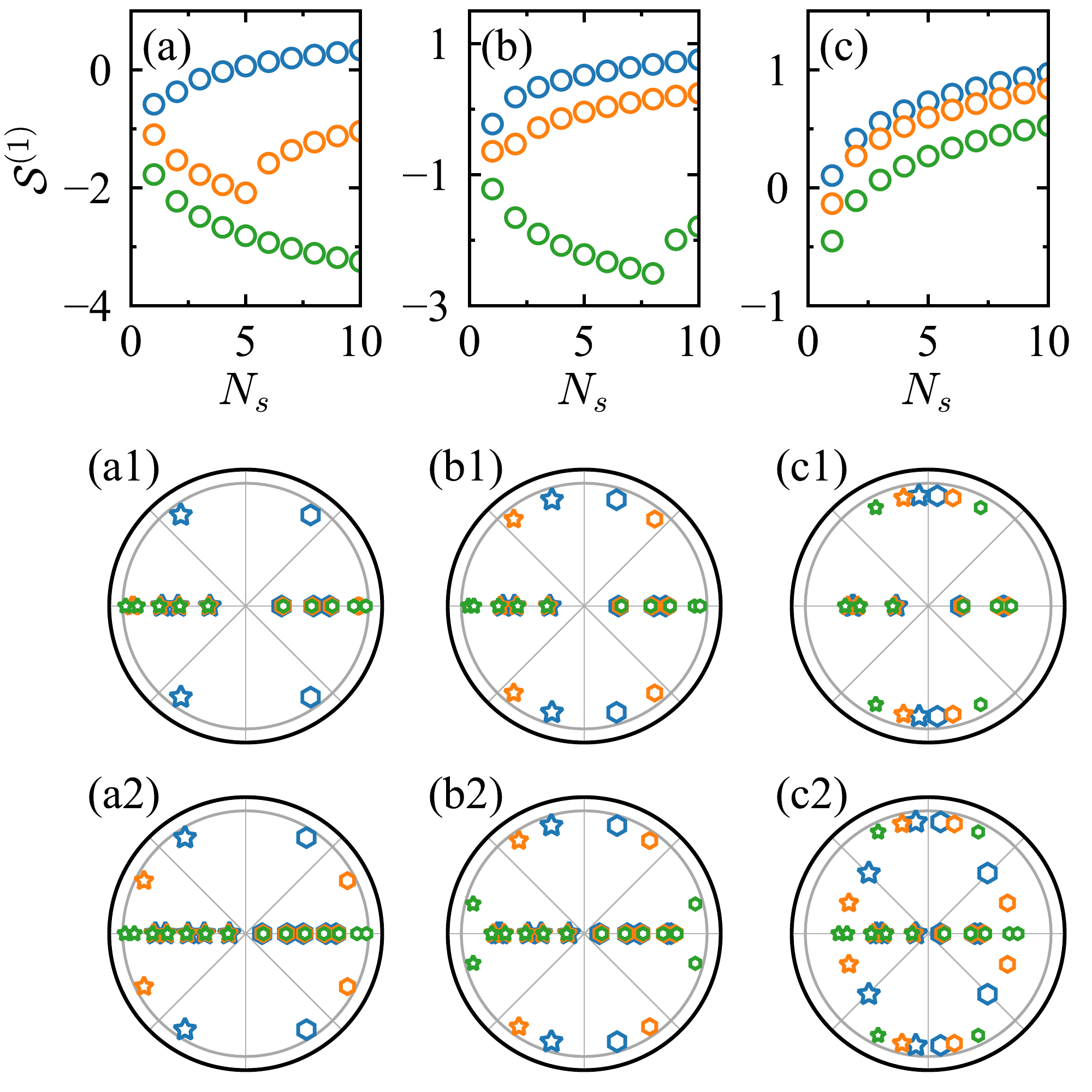}
\caption{
(a-c) Spatial profiles of the von Neumann entanglement entropy ${\cal S}$ ({\color{myblue} $\scircle[0.8]$}, {\color{orange} $\scircle[0.8]$} and {\color{mygreen} $\scircle[0.8]$}) as a function of the size $N_{s}$ of subsystem-$s$.
(a1-c1) and (a2-c2) $2$D radar maps with a radius $r=|z|$ and an angle $\phi=\arg(z)$, for visualizing the spectrum $\{w^{\cal C}_{q}\}$ of the correlation matrix $\cal C$.
The set-$0$ $z=w^{\cal C}_{q}$ ({\color{myblue} $\openbigstar$}, {\color{orange} $\openbigstar$} and {\color{mygreen} $\openbigstar$}) comprises all modes with $\text{Re}(w^{\cal C}_{q}) < 1/2$.
For the set-$1$ ({\color{myblue} $\varhexagon$}, {\color{orange} $\varhexagon$} and {\color{mygreen} $\varhexagon$}) that includes the remaining modes, we assign $z=w^{\cal C}_{q} - 1$.
The display area is limited by a high bound $1$ (grey outer circle border) and a low bound $10^{-9}$ (origin) in the log-$10$ scale.
We investigate three system size: $N=128$ (a, a1, a2), $256$ (b, b1, b2) and $1024$ (c, c1, c2), with three typical parameters $u=0.19998$ (blue), $0.199995$ (orange) and $0.199999$ (green) while fixing $t_2=0.8$.
Moreover, in radar maps, we zoom in on two cases of distinct cutting lengths $N_s=5$ (a1-c1) and $10$ (a2-c2).
Here, we choose $t_1=1$ with PBC.
}\label{sfig4}
\end{figure}

In Sec.~VI, we demonstrated that the ${\cal P}_{s}{\cal T}_{s}$ symmetry might break when the subsystem-$s$ size $N_{s}$ is larger than a certain healing size $N_{h}$.
We now estimate the change in the von Neumann entanglement entropy (vNEE) for a pair of eigenstates in the spectrum $\{w^{\cal C}_{q}\}$ during the ${\cal P}_{s}{\cal T}_{s}$ symmetry breaking process.
In the quadratic model of fermions, like our model (1) in the main text, vNEE ${\cal S} = -\sum_{q} [w^{\cal C}_{q} \ln w^{\cal C}_{q} + (1-w^{\cal C}_{q}) \ln (1-w^{\cal C}_{q})]$~\cite{Peschel_2009}.
In the ${\cal P}_{s}{\cal T}_{s}$ unbroken region $N<N_{h}$, two real eigenvalues $r_1$ and $r_2$ give ${\cal S}_{12} = -r_1 \ln r_1 - r_2 \ln r_2 - (1-r_1) \ln (1-r_1) - (1-r_2) \ln (1-r_2)$.
However, once $N_s > N_{h}$, these eigenvalues become a complex conjugate pair $w^{\cal C}_{q} = r_3 \exp(\pm i\phi_{3})$.
Assuming that $1 - w^{\cal C}_{q} = r_4 \exp(\pm i\phi_{4})$ with $r_4 = \sqrt{1+r^2_3-2r_3 \cos \phi_3}$, $\cos \phi_4 = (1-r_3 \cos \phi_3) / r_4$ and $\sin \phi_4 = -r_3 \sin \phi_3 / r_4$, the total contribution is given by
\begin{eqnarray}
{\cal S}_{34} &=& -2 (r_3 \ln r_3) \cos\phi_3 + 2 r_3 \phi_3 \sin\phi_3\nonumber\\
& & -2 (r_4 \ln r_4) \cos\phi_4 + 2 r_4 \phi_4 \sin\phi_4.
\end{eqnarray}
In the vicinity of EP $N_{s}=N_{h}$, the smoothly connecting condition implies $2 r_3 \cos\phi_3 = r_1 + r_2$ and $2 r_4 \cos\phi_4 = 2 - r_1 - r_2$.
Therefore, we have $-2 (r_3 \ln r_3) \phi_3 > - r_1 \ln r_1 - r_2 \ln r_2$ and $-2 (r_4 \ln r_4) \phi_4 > - (1 - r_1) \ln (1 - r_1) - (1 - r_2) \ln (1 - r_2)$, which yields the conclusion that ${\cal S}_{12} < {\cal S}_{34}$.
Consequently, vNEE increases when a real pair becomes a complex conjugate one.

\begin{figure}
\includegraphics[width=\linewidth]{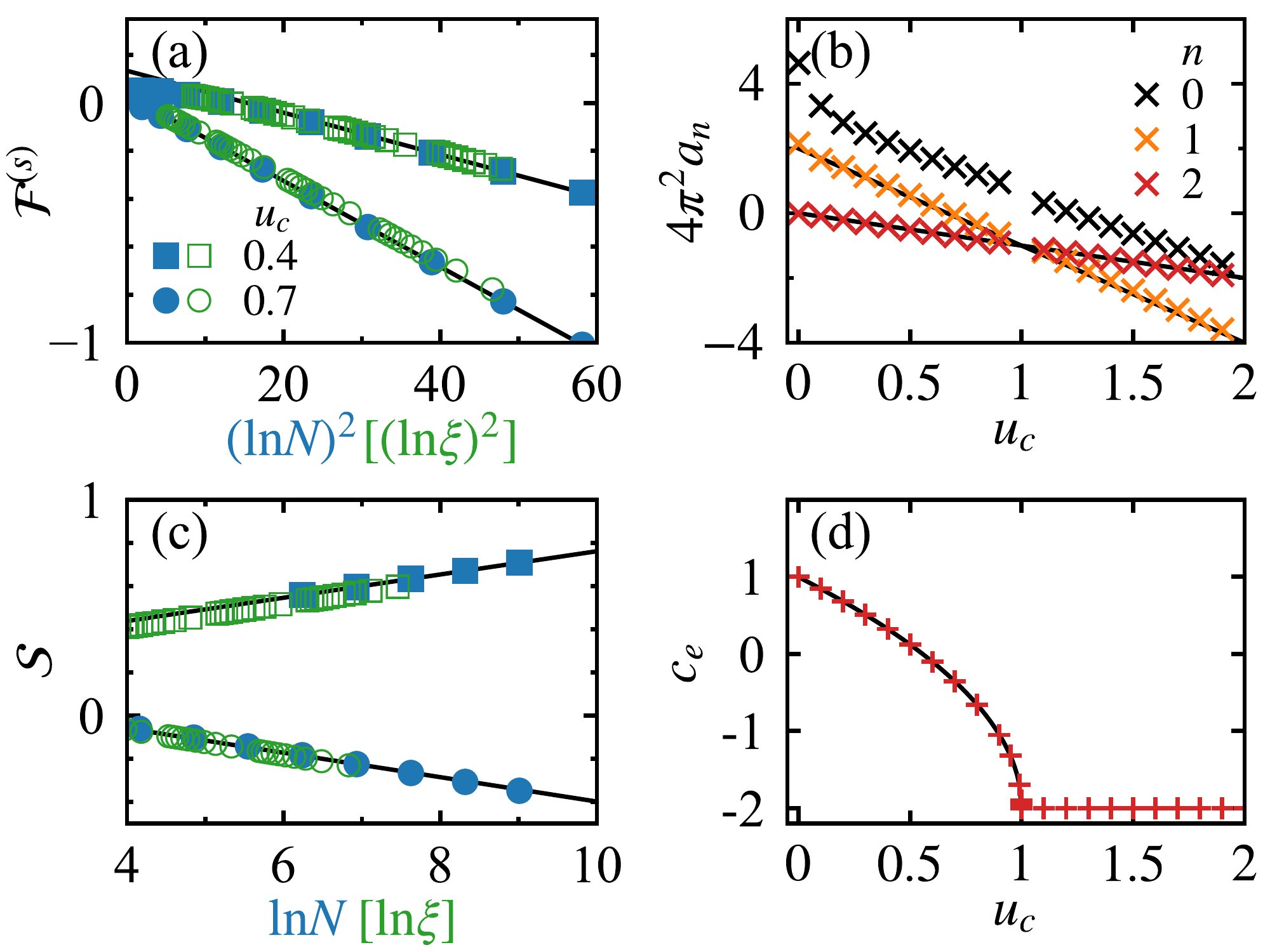}
\caption{
(a) The bipartite particle number fluctuations ${\cal F}^{(s)}$ as a function of $(\ln N)^2$ (blue) or $(\ln\xi)^2$ (green).
(b) The coefficients $a_n$ ({\color{black} $\times$}, {\color{orange} $\times$} and {\color{myred} $\times$}) in ${\cal F}^{(s)} = a_{2} (\ln N)^2 + a_{1} (\ln N) + a_{0}$ as a function of $u_{c}$.
(c) The bipartite von Neumann entanglement entropy as a function of $\ln N$ (blue) or $\ln\xi$ (green).
(d) The effective central charge $c_{e}$ ({\color{myred} $+$}) in the CFT prediction ${\cal S} \propto (c_{e}/6) \ln N$ as a function of $u_c$.
In both (a) and (c), we choose two typical parameters $u_c = 0.4$ ({\color{myblue} $\sblacksquare[0.8]$} and {\color{mygreen} $\ssquare[1.1]$}) and $0.7$ ({\color{myblue} $\sbullet[1.4]$} and {\color{mygreen} $\scircle[0.8]$}).
Meanwhile, black lines give the best fittings of ${\cal F}^{(s)}$ and ${\cal S}$, respectively. 
We choose $t_1=1$ with OBC.
}\label{sfig5}
\end{figure}

In Fig.~\ref{sfig4}, we examine the spatial profile of vNEE near EP $u_c = 0.2$ at $t_2=0.8$.
At $u=0.19998$ relatively far away from EP with a distance $\delta u = (u_c - u) = 2\times 10^{-5}$, the spectrum of the correlation matrix $\cal C$ displayed in radar maps exhibits a complex conjugate pair, as depicted in Fig.~\ref{sfig4}(a1-c1, a2-c2).
The resulting spatial profile of vNEE is convex, as shown in Fig.~\ref{sfig4}(a-c).
Closer to EP, specifically at $u=0.199999$ or $\delta u=10^{-6}$ with a larger correlation length, we observe that the spectrum is purely real for small $N=128$, but a complex conjugate pair emerges for larger $N=1024$.
Accordingly, the spatial profile of vNEE transforms from a concave to a convex function.
For an intermediate size $N=256$, a novel kink in Fig.~\ref{sfig4}(b) indicates the appearance of a complex conjugate pair.

\vspace{1cm}
\section{VIII.~Open boundary condition}\label{OBC}
For the open boundary condition (OBC) lacking an interface, we observe that BvNEE and BPNFs still behave as a polynomial function of either $\ln N$ or $\ln\xi$ in Fig.~\ref{sfig5}(a, c).
Moreover, coefficients $a_1$ and $a_2$ are half of the counterparts for PBC, as shown in Fig.~\ref{sfig5}(b).
This is consistent with the CFT prediction for Hermitian quantum criticality.
But, the curve of $a_0$ has an exceptional pace near $u_c = 0$, lacking a theoretical explanation yet.

\bibliography{refSM}